\documentclass[fleqn,10pt]{wlscirep}
\usepackage[utf8]{inputenc}
\usepackage[T1]{fontenc}

\usepackage{bm}
\usepackage{amsmath,amssymb}
\usepackage{cite}
\usepackage{xcolor}

\usepackage{comment}

\definecolor{dark_green}{rgb}{0.0, 0.5, 0.0}

\newcommand{\chng}{\color{black}}

\newcommand{\M}{{\cal M}}
\newcommand{\W}{{\cal W}}
\newcommand{\R}{{\cal R}}
\newcommand{\K}{{\cal K}}
\newcommand{\cO}{{\cal O}}
\newcommand{\ds}{\displaystyle}

\newcommand{\onlinecite}[1]{\hspace{-1 ex} \nocite{#1}\citenum{#1}} 

\title{Acoustic frequency combs using gas bubble cluster oscillations
  in liquids: \chng{a proof of concept}}

\author[1]{\chng Bui Quoc Huy Nguyen}
\author[1,*]{\chng Ivan S.~Maksymov}
\author[2]{Sergey A.~Suslov}
\affil[1]{Optical Sciences Centre, Swinburne University of
  Technology, Hawthorn, Victoria 3122, Australia\looseness=-1}
\affil[2]{Department of Mathematics, Swinburne University of
  Technology, Hawthorn, Victoria 3122, Australia\looseness=-1}

\affil[*]{imaksymov@swin.edu.au}

\begin{abstract}
We propose a new approach to the generation of acoustic
frequency combs (AFC)---signals with spectra containing equidistant
coherent peaks. AFCs are essential for a number of sensing 
and measurement applications, where the established technology
of optical frequency combs suffers from fundamental physical
limitations. Our proof-of-principle experiments demonstrate that
nonlinear oscillations of a gas bubble cluster in water insonated by a
low-pressure single-frequency ultrasound wave produce signals 
with spectra consisting of equally spaced peaks originating from
the interaction of the driving ultrasound wave with the
response of the bubble cluster at its natural frequency. The
so-generated AFC posses essential characteristics of optical frequency
combs and thus, similar to their optical counterparts, can be used to
measure various physical, chemical and biological quantities.
\end{abstract}

\begin{document}

\flushbottom
\maketitle
%
%
\thispagestyle{empty}

\section*{Introduction}

Optical frequency combs---optical spectra composed
of equidistant narrow peaks---enable precision
measurement in both fundamental and applied contexts
\cite{Pic19, For19, Hug12, Li19, Li20, Ste20}.
An optical frequency comb acts as a spectrum
synthesizer that enables the precise transfer of phase and frequency
information from a stabilised reference to optical signals. The
so-generated signals can be used, for example, to obtain the spectral
response of a gas or liquid sample due to linear or nonlinear
absorption of light by the medium \cite{Tho08}. One can also
accurately measure distances by passing an optical frequency comb
signal through an interferometer and then analysing the resulting
interference pattern, which is beneficial for the fields of satellite
positioning and material science \cite{Ber12}. 
\begin{figure}[t]
  \centerline{
    \includegraphics[width=10cm]{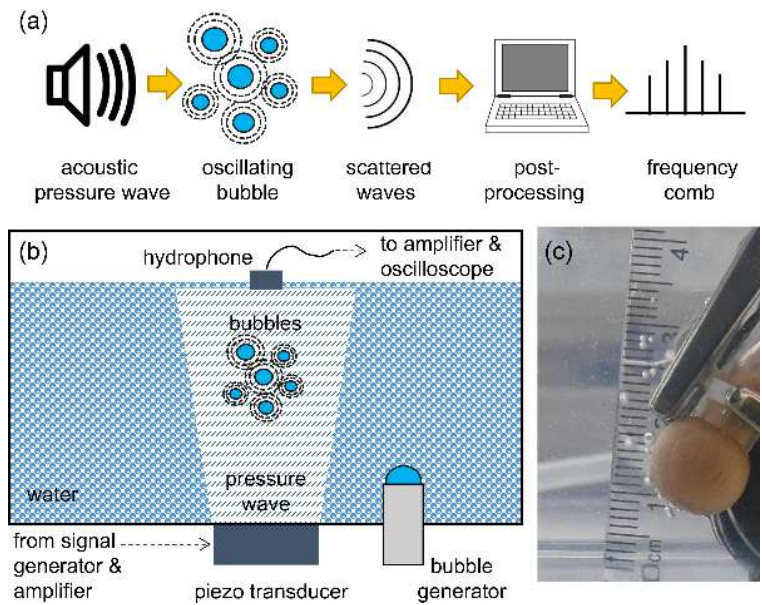}}
  \caption{(a)~Schematic diagram of the suggested AFC generation. The
    oscillations of the bubble cluster are driven by a 
    single-frequency ultrasound pressure wave. Acoustic waves
    scattered by the bubble cluster are recorded and post-processed to 
    obtain a spectrum consisting of the equidistant
    peaks. (b)~Schematic of the experimental setup. Bubbles are
    created in a stainless steel tank using a bubble generator. The
    driving pressure wave is emitted by an ultrasonic
    transducer. Waves scattered by the bubble are detected by a
    hydrophone.
    (c)~Photograph of typical gas bubbles emitted by the bubble
    generator in a water tank with transparent walls at otherwise 
    identical experimental conditions to those in the stainless steel 
    tank. The diffuser of the bubble generator and other elements 
    of the setup can be seen.}
  \label{fig:fig1}
\end{figure}

However, using optical frequency combs is not always possible
because of a number of fundamental and technical limitations.
For example, in liquid samples such as biological fluids
light can be strongly reflected and absorbed by the medium.
Photoacoustic frequency comb spectroscopy may help to
partially resolve these problems \cite{Wil20, Fri20},
because this technique exploits absorption of light and
concomitant generation of acoustic waves that carry
information about the absorption strength. Yet, more versatile and
technologically simple approaches are still required.   

Similar to optical frequency combs, acoustic (phononic)
frequency combs (AFC)---purely acoustic signals with spectra
containing equidistant coherent peaks---exploit the
ability of acoustic waves to provide precision information
about the medium in which they propagate
\cite{Cao14, Xio16, Cao16, Gan17, Wu19, Mak19, Gor20}. In contrast
to light, acoustic waves can propagate in water and opaque
liquids over long distances, which underpins many
acoustics-based technologies including sonar, underwater
communication and sensing \cite{Wu19} and marine biology \cite{Uri83}.  
Yet, even though AFCs have already been used to
accurately measure distances between underwater objects \cite{Wu19},
research on them remains under-established. The development of new types
of acoustic combs is needed for sensing and imaging systems
\cite{Cao14, Gan17}, in particular, biomedical imaging \cite{Fai77,
  Miw84, Mak19}.

In this work, we demonstrate the possibility of the AFC generation
using a gas bubble cluster nonlinearly oscillating in water \cite{Bre95},
when it is driven by a single-frequency ultrasound wave
[Fig.~\ref{fig:fig1}(a)]. Unlike in the scenario of an optical
frequency comb generation using a high-power laser light and
exploiting fundamentally weak nonlinear-optical effects \cite{Mak19},
we show that the application of low-pressure harmonic signals can
trigger a strong nonlinear response of the cluster resulting in the
generation of multiple ultraharmonic frequency peaks. The interaction
with {\chng stochastic} noise-induced bubble cluster oscillations at its natural
frequency, which is typically much lower than that of a driving
ultrasound, results in the amplitude modulation of the the bubble
cluster response and the appearance of sidebands around the main
peaks. {\chng This process is thresholdless and therefore the
  sidebands can always be experimentally observed when their amplitude
  is higher than the noise floor.}

Our current findings contribute to further development of an emergent
field of AFC generation \cite{Cao14, Gan17, Wu19, Mak19, Gor20}. 
They also extend our previous observation of frequency combs
originating from the onset of Faraday waves in vertically vibrated
liquid drops \cite{Mak19_PRE, Mak19_SPIE}. However, in that system
the spacing between the peaks of the comb was only 20--40\,Hz.
Whereas frequency combs with a Hz-range spacing can find certain 
applications \cite{Gor20}, in the gas bubble system investigated
in the present work we use the high-kHz range that can potentially be
extended to the high-MHz range \cite{Mak18}. This opens up 
opportunities for using acoustic combs instead of optical ones or in
addition to them in a number of practical situations where operation at
higher frequencies may be required \cite{Mak19}.

\section*{Methods}
\subsection*{Experiment}
\label{sec:experiment}
Our experimental setup shown in Fig.~\ref{fig:fig1}(b)
consists of a 1.5\,L thin-walled stainless steel tank filled 
with distilled degassed water maintained at room temperature. 
A generic piezoceramic disc transducer with the measured 
resonance frequency of $42.6\pm 0.3$\,kHz is glued 
to the bottom wall of the tank from outside. The entire apparatus is 
assembled on a customised vibration damping support. {\chng The
  laboratory room is located in a single-level standalone building
  with an approximately $50$\,cm thick concrete floor. The building is
  located several kilometres away from railways, highways and building
  sites, where heavy machinery and equipment are used. Together with
  electric shielding provided by metalled walls, this ensures that
  external vibrations and interference with high-voltage equipment do
  not affect results of our measurements.}

The piezo transducer is driven by a digital tone generator (Rigol 
DG-1022Z, China) connected to a broadband power amplifier 
(Bosch Plena LBB1906/10, Germany). The signal is fed to the piezo-disc 
via an electrical impedance-matching circuit featuring a customised 
adjustable mH-range inductor coil (Scientific, Australia) connected 
in series with the piezo-ceramic transducer. Effectively, the 
piezo-ceramic disc behaves as a capacitor that draws little 
current from the amplifier but requires high voltage that 
in our case is produced by resonance tuning of the $LC$ circuit on 
the frequency of interest. Hence, in our measurements we 
fix the frequency of the driving ultrasound wave and change its 
pressure amplitude because this does not require re-tuning the 
inductor coil.  

The hydrophone is based on a small piezoceramic disc 
(type PIC155, PICeramic, Germany). Electric signals produced 
by the piezo disc are first amplified using a broadband 
voltage amplifier (BWD 603B, Australia) with the frequency 
response from 0 to 100\,kHz. Then the signal is sent 
to a digital oscilloscope (Rigol, DS-1202ZE, China) 
controlled via a laptop computer. 

We use a customised bubble generator consisting of an air 
pump connected to a silicone tubing terminating 
in a diffuser made of a piece of porous material. The generator 
produces several single bubbles per second with the radius 
of $1.0\pm 0.5$\,mm. The size of the bubbles was estimated 
using high-speed digital video camera records 
[see Fig.~\ref{fig:fig1}(c)]. 

The large bubbles rise under the effect of buoyancy. Some of them
become trapped in the middle of the water layer due to the primary 
Bjerknes force of the ultrasound pressure wave \cite{Bre95}. As a 
result, a cluster of bubbles is formed. We record scattered
pressure signals produced by the bubble clusters using the
hydrophone and then process the measured time-domain signals
in the Octave software to obtain spectral information.

By using high-speed imaging we estimate that on average the radius of
a bubble cluster is $R_c=20\pm5$\,mm and the air fraction (the ratio
of the total volume of bubbles in a cluster to the volume of a region
occupied by the bubble cluster) inside it \cite{dAg88, dAg89, Mae19}
is $\chi\approx0.015$. This implies that in our case the product
$\sqrt{\chi}R_c$ is larger than the radius of the largest bubble in
the cluster. According to Eq.~(\ref{eq:eq6}), this means that the
detected frequency $f_c$ should be smaller than the natural frequency
of individual bubbles, which is indeed confirmed by our experimental
observations.

\subsection*{Model}
\label{sec:model}
Modelling bubble clusters is a challenging task given that their
geometry varies from experiment to experiment and with
time. Therefore, models considering a cluster as a single equivalent 
bubble of the size larger than that of constituent bubbles are
frequently used \cite{dAg88, dAg89, Hwa00}, especially when of
interest is the natural oscillation frequency of the cluster as a
whole, which is the case in the current work. However, when
doing so, one needs to keep in mind that the ultrasound energy
absorption and scattering characteristics of a cluster may differ from
those of an equivalent single bubble \cite{Kaf00, dAg88}. The scattering
($\sigma_{scat}$) and absorption ($\sigma_{abs}$) cross-sections of a
single bubble placed in the field of an incident plane ultrasound wave
are defined as the ratios of the scattered and absorbed powers,
respectively, to the power of the incident wave \cite{Bre03,
  Ain11}. The extinction cross-section
$\sigma_{ext}=\sigma_{scat}+\sigma_{abs}$ characterises the incident
wave energy loss due to its absorption and scattering by the
bubble. Generally speaking, a large gas bubble behaves as a strong
acoustic scatterer with $\sigma_{scat}$ proportional to the square of 
the bubble radius \cite{Kaf00}. Because $\sigma_{abs}$ also scales 
with the square of the radius \cite{Bre03, Ain11}, larger 
bubbles have larger $\sigma_{ext}$. However, the scattering 
cross-section of a bubble cluster is also proportional to the air 
fraction inside the cluster: $\sigma_{scat\,c}=\chi R_c^2$. 
Because in our case $\chi\approx0.015$, the absorption and scattering
by the bubble cluster detected in experiments are smaller than those
by a single equivalent bubble. 

The excitation of a bubble cluster with a single-frequency signal is
also known to result in a stronger nonlinear generation of
ultraharmonics \cite{Rud06, Ain11} compared with the case of a single
bubble. This is also observed in our experiments. However, these
features are inconsequential in the context of AFC generation, which
is the main focus of the current work. Therefore, we rely on the
results obtained using an equivalent single bubble model discussed
below to explain the experimentally detected acoustic response
spectra.

The accepted model of nonlinear oscillations of a single 
spherical bubble in water is given by the Keller-Miksis (KM) 
equation \cite{Kel80}. It takes into account the decay of bubble 
oscillations due to viscous dissipation and fluid
compressibility. However, in this work we investigate  
millimetre-sized gas bubbles in water oscillating at 20--100\,kHz
frequencies and driven by low pressure waves with amplitude 
of up to 25\,kPa. We established that in this regime 
the terms of the KM equation accounting for acoustic losses 
are negligible. Thus, in the following we omit these 
terms thereby effectively reducing the KM model to the classical 
Rayleigh-Plesset (RP) equation \cite{Ray17, Ple49}: 
{\chng
\begin{equation}
  \label{eq:eq1}
  R\frac{d^2R}{dt^2}+\frac{3}{2}\left(\frac{dR}{dt}\right)^2 
  =\frac{1}{\rho}\left(P\left(R,\frac{dR}{dt}\right)-P_\infty(t)\right)\,,\quad
  R(0)=R_0+\tilde R_0\,,\quad\frac{d}{dt}R(0)=V\,,
\end{equation}
where $\tilde R$ is the initial deviation of the bubble radius from
the equilibrium value $R_0$, $V$ is the initial speed of the bubble wall,
\begin{equation}
  \label{eq:eq2}
  P\left(R,\frac{dR}{dt}\right)=\left(P_0-P_v+\frac{2\sigma}{R_0}\right) 
  \left(\frac{R_0}{R}\right)^{3\kappa}
  -\frac{4\mu}{R}\frac{dR}{dt}-\frac{2\sigma}{R}
\end{equation}
and the expression $P_\infty(t)=P_0-P_v+\alpha\sin(\omega^* t)$ with 
$\omega^*=2\pi f_0$ represents the periodically varied pressure in the 
liquid far from the bubble. The parameters $R(t)$, $\mu$, $\rho$, 
$\kappa$, $\sigma$, $\alpha$, and $f_0$ denote, respectively, the 
instantaneous bubble radius}, the dynamic 
viscosity and the density of the liquid, the polytropic exponent of a 
gas entrapped in the bubble, the surface tension of a gas-liquid 
interface and the amplitude and the frequency of a driving
ultrasound wave. The diffusion of the gas through the bubble surface
is neglected. 

In our model oscillations of the bubble are not affected by fluid
compressibility, and we can express the acoustic power scattered by
the bubble into the far-field zone as \cite{Bre95}
\begin{equation}
  \label{eq:eq3}
  P_{scat}(R,t) = \frac{\rho R}{h}\left(R\frac{d^2R}{dt^2}
      +2\left(\frac{dR}{dt}\right)^2\right)\,,
\end{equation}
where $h\gg R_0$ is the distance from the centre of the bubble. The 
natural frequency of the bubble is \cite{Bre95}
{\chng
  \begin{eqnarray}
    &f_{nat} = \frac{1}{2\pi\sqrt{\rho} R_0}\sqrt{3\kappa 
      \left(P_0-P_v+\frac{2\sigma}{R_0}\right)-\frac{2\sigma}{R_0}
      -\frac{4\mu^2}{\rho R_0^2}}
      \approx f_M\left(1
      +\frac{(3\kappa-1)\sigma}{3\kappa R_0(P_0-P_v)}
      -{\chng\frac{2\mu^2}{3\kappa\rho R_0^2(P_0-P_v)}}
      \right)\,,\quad
      f_M=\frac{\sqrt{3\kappa\left(P_0-P_v\right)}}
      {2\pi R_0\sqrt{\rho}} &\label{eq:eq5}
  \end{eqnarray}
  where $f_M$} is the well-known Minnaert frequency \cite{Min33}.
We use the following fluid parameters corresponding to water 
at $20^\circ$\,C: $\mu=10^{-3}$\,kg\,m/s, 
$\sigma=7.25\times10^{-2}$\,N/m, $\rho=10^3$\,kg/m$^3$
and $P_v=2330$\,Pa. In our computations we take the air 
pressure in a stationary bubble to be $P_0=10^5$\,Pa  and the 
polytropic exponent of air to be $\kappa=4/3$ \cite{Cru83, 
  Pay05}. For mm-sized air bubbles in water {\chng the surface tension 
  and viscous terms in parenthesis in Eq.~(\ref{eq:eq5}) are of the order of
  $10^{-4}$ and $10^{-9}$, respectively. Therefore, both the increase
  of the bubble natural frequency due to the influence of the surface
  tension and its decrease due to viscous effects can be neglected.}
The natural frequency of a bubble cluster is given by
\begin{equation}
  \label{eq:eq6}
  f_c\approx \frac{f_M}{\sqrt{\chi}}\frac{R_0}{R_c}\,,
\end{equation} 
where $R_c$ is the radius of the bubble cluster and $\chi$ is 
the air fraction in the liquid \cite{Hwa00}. In our experiments, we
established that $\sqrt{\chi}R_c > R_0$. 

Equation~(\ref{eq:eq1}) was solved numerically using an explicit 
Runge-Kutta method \cite{Dor80} implemented in a standard 
subroutine \texttt{ode45} in the Octave software. The numerical
solution was used to obtain the acoustic scattering spectra calculated
using Eq.~(\ref{eq:eq3}). In the solver configurations, the numerical
values of the absolute and relative error tolerances were set to
machine accuracy. {\chng However, to identify the main
  characteristics of nonlinear bubble oscillations relevant to the AFC
  generation, we present an asymptotic analysis of
  Eq.~(\ref{eq:eq1}) next.

  Firstly, Eq.~(\ref{eq:eq1}) is re-written in the non-dimensional
  form, where we use the equilibrium radius $R_0$ and
  $t_0=1/\omega^*$ as the length and time scales,
  respectively, to introduce non-dimensional quantities
  $r=R(t)/R_0$ and $\tau^*= \omega^*t$. Substituting these
  into Eq.~(\ref{eq:eq1}), we obtain
  {\chng
    \begin{equation} 
      rr''+\frac{3}{2}(r')^2=\frac{\M+\W}{r^\K}-\M-\frac{\W}{r}-\R\frac{r'}{r}
      +\M_e\sin\tau^*\,,\quad
      r(0)=1+\frac{\tilde R_0}{R_0}\,,\quad
      r'(0)=\frac{V}{\omega^*R_0}\,,\label{NDRPE}
    \end{equation}
  where prime} denotes differentiation with respect to $\tau^*$,
  $\ds\M=\frac{P_0-P_v}{\rho{\omega^*}^2R^2_0}$, 
  $\ds\W=\frac{2\sigma}{\rho{\omega^*}^2R^3_0}$,
  $\ds\R=\frac{4\mu}{\rho{\omega^*} R^2_0}$,
  $\ds\M_e=\frac{\alpha}{\rho{\omega^*}^2R^2_0}$ and
  $\K=3\kappa$. As discussed in ref.~\onlinecite{Sus12},
  parameter $\M$ represents elastic properties of the gas and its
  compressibility, $\W$ and $\R$ can be treated as inverse Weber and
  Reynolds numbers, characterising the surface tension and viscous
  dissipation effects, respectively, and $\M_e$ is the measure of the
  ultrasound forcing. For the conditions of our experiment $\K=4$ and
  the maximum values of other parameters do not exceed
  $\M=1.8\times10^{-3}$, $\W=1.8\times10^{-6}$,
  $\R=1.2\times10^{-5}$ and $\M_e=2.1\times10^{-4}$. As shown in
  ref.~\onlinecite{Sus12}, the attenuation of bubble oscillations
  due to viscous dissipation is proportional to $\exp(-\R\tau'/2)$ and
  thus is negligible (consequently, we set $\R=0$ in what
  follows). This suggests that, in the context of the current study,
  forced bubble oscillations can be assumed perfectly periodic
  provided that the driving ultrasound frequency is much higher than
  any of the bubble resonance frequencies (see  below). This enables
  us to make an analytic progress by employing Poincar\'e-Lindstedt
  method, which is algebraically simpler than Bogolyubov-Krylov
  \cite{Prosperetti74} or multiple scales \cite{Francescutto83,
    Francescutto84} methods traditionally used to account for various
  transient processes in bubble dynamics. 
  
  We introduce a stretched time variable
  $\tau=\omega\tau^*=\omega\omega^*t$, where $\omega$ is the
  non-dimensional frequency of nonlinear bubble oscillations that
  depends on their amplitude. The estimation of physical parameters
  given above suggests that ultrasound forcing applied to a bubble is
  relatively weak and, subsequently, it is not expected to cause large
  amplitude oscillations of a bubble. Therefore, we look for an
  asymptotic solution of  Eq.~(\ref{NDRPE}) in the form
  \begin{eqnarray}
    r&=&1+\epsilon
         r_1(\tau)+\epsilon^2r_2(\tau)+\epsilon^3r_3(\tau)+\ldots\,,
         \label{r_taylor}\\
    \omega&=&\omega_0+\epsilon\omega_1+\epsilon^2\omega_2+
              \ldots\,\label{omega_taylor},
  \end{eqnarray}
  where $\epsilon$ is a formal parameter introduced to distinguish
  between various terms in the asymptotic series (it is to be set to
  unity once the series is developed). We also set $\W=0$ given that, as
  estimated above, the influence of surface tension in our experimental
  setup is so weak that it cannot be detected within the measurement
  accuracy. Introducing such a simplification in the mathematical model
  does not have any effect on the functional form of solutions either
  but makes the expressions for various solution coefficients much
  shorter and easier to interpret. Choosing
  $\omega_0^2=\K(\M+\W)-\W=\K\M$, which corresponds to Minnaert  
  frequency given by Eq.~(\ref{eq:eq5}), substituting Eq.~(\ref{r_taylor}) and
  Eq.~(\ref{omega_taylor}) into Eq.~(\ref{NDRPE}) and collecting terms at
  various orders of $\epsilon$, we obtain the following equations (up to
  $\cO(\epsilon^3)$)
  \begin{eqnarray}
    \ddot{r}_1+r_1&=& p\sin(\Omega\tau)\,,\label{eqeps1}\\
    \ddot{r}_2+r_2&=&\frac{1}{2}(1+\K)r_1^2
                      -\frac{3}{2}\dot{r}_1^2 - r_1\ddot{r}_1
                      -2\frac{\omega_1}{\omega_0}\ddot{r}_1\,,\label{eqeps2}\\
    \ddot{r}_3+r_3&=&(1+\K)r_1r_2-\frac{1}{3}\left(1+\frac{1}{2}\K(\K+3)\right)r_1^3
                      +\frac{3\omega_1\dot{r}_1^2}{\omega_0}+3\dot{r}_1\dot{r}_2
                      +\left(\frac{\omega_1^2}{\omega_0^2}
                      +\frac{2\omega_2}{\omega_0}\right)\ddot{r}_1
                      + \ddot{r}_1r_2+\frac{2\omega_1}{\omega_0}r_1\ddot{r}_1\,,
                      \label{eqeps3}
  \end{eqnarray}
  {\chng where dot denotes differentiation with respect to $\tau$ and we
    set $(\M_e/\omega_0^2)\sin\tau^*=\epsilon p\sin(\Omega\tau)$,
    $\Omega=1/\omega\gg1$. We also write the random initial conditions
    as $r(0)=1+\epsilon a$ and $\dot{r}(0)=\epsilon b$, where $\epsilon
    a=\tilde R_0/R_0$ and $\epsilon b=(V\Omega)/(\omega^*R_0)$.} Then 
  \begin{equation}
    r_1(0)=a\,,\ r_2(0)=r_3(0)=0\quad\mbox{and}\quad
    \dot{r}_1(0)=b\,,\ \dot{r}_2(0)=\dot{r}_3(0)=0\,.
  \end{equation}
  The solution $r_1(\tau)$ is 
  \begin{equation}
    r_1=a\cos\tau+(b+D\Omega)\sin\tau-D\sin(\Omega\tau)
    =R_1\sin(\tau+\phi_1)-D\sin(\Omega\tau)\,,\label{eqr1a} 	
  \end{equation}
  where {\chng $\ds D=\frac{p}{\Omega^2-1}$,}
  $R_1=\sqrt{a^2+(b+D\Omega)^2}$ and 
  $\phi_1=\tan^{-1}{\dfrac{a}{b+D\Omega}}$. This solution is valid away
  from the main resonance, that is for $\Omega\ne1$, which is the case in
  the considered application since the bubble natural frequency is much
  smaller than that of a driving ultrasound wave and thus
  $\Omega\gg1$. In particular, in our experiments
  $\Omega\approx1/\omega_0\approx11.6$. The relevant physical
  conclusion that follows from Eq.~(\ref{eqr1a}) is that unless the
  initial conditions are chosen in a very specific way ($a=0$ and
  $b=-D\omega$), the leading order bubble response will always contain
  two frequencies: the bubble's natural frequency and the driving
  ultrasound frequency. It is conceivable that in very careful
  experiments with a single bubble the excitation of its oscillations
  at the natural frequency would not occur, or they could decay due to
  dissipation if the observation time is sufficiently long. Then the
  generation of the bubble-based AFC would fail. However, this
  scenario is improbable when a bubble cluster is used, where
  neighbouring bubbles acoustically interact with each other so that
  initial conditions $a$ and $b$ for each individual bubble remain
  essentially arbitrary. {\chng (Indeed, in our earlier experiments with
    a single large bubble the number of cases, where the oscillations
    at the natural frequency were not observed, was statistically
    insignificant and AFCs have been always observed in our
    experiments with bubble clusters.)} 

  The second relevant conclusion that solution Eq.~(\ref{eqr1a})
  offers is that in the non-resonant conditions, which are of interest
  here, the amplitude of both forced and natural bubble oscillation is
  proportional to that of the driving field. At the same time, the
  amplitude of forced oscillations is inversely proportional to the
  square of the driving frequency, while that of the natural
  oscillations to its first power. This offers a practical opportunity
  of controlling the power spectrum of the so-generated AFC by
  optimising the choice of the $(\alpha,\omega')$ characteristics of
  insonification.

  To avoid an aperiodic secular term proportional to $\tau$ in the
  solution of the second-order equation we must eliminate the last
  term in the right-hand side of Eq.~(\ref{eqeps2}) by setting
  $\omega_1=0$. Then solving Eq.~(\ref{eqeps2}) with $r_1(\tau)$ given
  by Eq.~(\ref{eqr1a}) we obtain
  \begin{eqnarray}
    r_2&=&C+c_1\cos\tau+c_2\sin\tau-A\cos(2\tau)-B\sin(2\tau)
           +E\cos(2\Omega\tau)\nonumber\\ 
       &&+F
          \cos(\Omega\tau)[(b+D\Omega)\cos\tau-a\sin\tau] 
          +G \sin(\Omega\tau)[a\cos\tau+(b+D\Omega)\sin\tau]
          \nonumber\\
       &=&C+R_2\sin(\tau+\phi_2)-R_3\sin(2\tau+\phi_3)
           +E\cos(2\Omega\tau)
           +\frac{a}{2}\left[(G+F)\sin[(\Omega-1)\tau]
           +(G-F)\sin[(\Omega+1)\tau]\right]\nonumber\\
       &&+\frac{b+D\Omega}{2}
          \left[(G+F)\cos[(\Omega-1)\tau]-(G-F)\cos[(\Omega+1)\tau]\right]\,,
          \label{eqr2b}
  \end{eqnarray}
  where
  \begin{eqnarray*}
    &c_1=A-C-E-F(b+D\Omega)\,,\quad c_2=2B+a(F-G\Omega)\,,&\\
    &\ds A=\frac{1}{12}[a^2-(b+D\Omega)^2](6+\K)\,,\quad
      B=\frac{a}{6} (b+D\Omega)(6+\K)\,,&\\
    &\ds C=\frac{\K}{4}\left[a^2+(b+D\Omega)^2+D^2\right]
      +\frac{D^2}{4}(1-\Omega^2)\,,\quad
      E=\frac{D^2}{4}\frac{5\Omega^2+\K+1}{4\Omega^2-1}\,,&\\
    &\ds F=-\frac{D}{\Omega}\left(1-\frac{2\K}{\Omega^2-4}\right)\,,\quad
      G=D\left(1+\frac{\K}{\Omega^2-4}\right)\,,&\\
    &R_2=\sqrt{c_1^2+c_2^2}\,,\quad\phi_2=\tan^{-1}{\dfrac{c_1}{c_2}}\,,\quad
      R_3=\sqrt{A^2+B^2}\,,\quad\phi_3=\tan^{-1}{\dfrac{A}{B}}\,.&
  \end{eqnarray*}
  A number of relevant conclusions can be made from the expression for
  $r_2$. Firstly, due to the nonlinearity the second harmonics of both
  natural (2) and forcing ($2\Omega$) frequencies appear in addition to
  the fundamental harmonics (with non-dimensional frequencies 1 and
  $\Omega$) and their amplitudes (coefficients $A$, $B$ and $E$) are
  proportional to the squares of the initial conditions and the forcing
  amplitude. This means that if the forcing and random noise in the
  system are small, the magnitudes of the spectral lines at such double
  frequencies are smaller than those of base frequencies. Secondly, the
  amplitudes $A$ and $B$ of double natural frequency components decay as
  $\Omega^{-2}$ with the forcing frequency while the $2\Omega$
  ultraharmonic decays as $\Omega^{-4}$, see coefficient $E$. Therefore,
  at high-frequency forcing the ultraharmonics of forcing decay faster
  than those of natural frequency and this fact can be used to balance
  the AFC peaks centred at $n\Omega$, $n=2,3,\ldots$. Thirdly, the time
  average of the bubble radius deviates from $r=1$: since $\K>1$
  coefficient $C$ is positively defined reflecting the known fact
  that a nonlinearly oscillating bubble stays in an inflated state for
  the most of the oscillation period with short-lasting rapid
  contraction/restoration stages. Fourthly, the solution coefficients 
  become singular at forcing frequencies $\Omega=1/2$ and 2, which
  indicates the possibility of ultra- and subharmonic resonances when
  the current solution would break down. This is not of a concern in the
  AFC context though because here $\Omega\gg1$ by design.
  
  The asymptotic expansion procedure can be routinely continued to
  higher orders revealing the possibility of further ultra- and
  subharmonic resonances at $\Omega=n$ and $\Omega=1/n$ and amplitude
  modulations with $\Omega\pm n$, $n=1,2,3,\ldots$. However, here we do
  not pursue this any further and only note that the elimination of
  secular terms at order $\epsilon^3$ requires that
  \begin{eqnarray}
    \omega_2&=&\frac{\omega_0}{8}
                \left[D^2\left(\frac{2\K^2}{\Omega_0^2-4}+\K\Omega_0^2+2\right)
                -\frac{a^2+(b+D\Omega_0)^2}{6}\left(2\K^2-3\K-6\right)\right]
                \label{om2a}\\
            &=&\frac{\omega_0}{4}
                \left[D^2\left(\frac{16}{\Omega_0^2-4}+2\Omega_0^2+1\right)
                -\frac{7}{6}\left(a^2+(b+D\Omega_0)^2\right)\right]\,,\quad
                \Omega_0=\frac{1}{\omega_0}\,.\label{om2b}
  \end{eqnarray}
  This expression demonstrates that in the noisy conditions (when $a$
  and/or $b$ are non-zero) the frequency of nonlinear bubble
  oscillations decreases with the square of the initial conditions (the
  second term in the brackets of Eq.~(\ref{om2b}) is negative),
  which is a well established fact \cite{Prosperetti74, Francescutto83,
    Francescutto84}. {\chng However, the application of high-frequency forcing,
    when $\Omega\approx\Omega_0=1/\omega_0$ and
    $D\Omega^2\approx p=\M_e/\omega_0^2$,
    tends to partially compensate for such a noise-induced frequency
    reduction (the first term in Eq.~(\ref{om2b}) is positive for large
    $\Omega$). In terms of physical parameters the relative variation of
    the bubble natural oscillation frequency in an ultrasound field in
    this limit is}
  \begin{equation}
    \frac{\Delta f}{f_M}\approx\frac{\omega_2}{\omega_0}
    {\chng \approx\frac{5}{24}\left(\frac{\M_e}{\omega_0}\right)^2
      -\frac{7}{24}\left(a^2+b^2+2b\frac{\M_e}{\omega_0}\right)\,,
      \quad\frac{\M_e}{\omega_0}=\frac{\alpha}{\rho\omega_0{\omega^*}^2R_0^2}
      =\frac{\alpha}{4\pi^2\rho f_Mf_0R_0^2}}\,.\label{om2om0}
  \end{equation}
  {\chng The downshift of natural frequency of the bubble oscillation
    occurs when the noise level exceeds $\ds (a,b)\gtrsim(\M_e/\omega_0)$,
    which is $\sim10^{-3}$ in our experiments (i.e. when the deviation 
    of the bubble radius from the equilibrium value caused by a random
    noise exceeds 0.1\%). In this case, the frequency upshift due to ultrasound
    forcing is negligible and by measuring the downshift of the
    bubble's natural frequency one can use Eq.~(\ref{om2om0}) to estimate
    the average intensity $a^2+b^2$ of natural bubble oscillations.
    
    Finally, we set $\varepsilon=1$ and write the leading terms in the
    non-dimensional expression $p_{scat}=(P_{scat}h)/(\alpha R_0)$ for the
    acoustic power scattered by a bubble (see Eq.~(\ref{eq:eq3})) as}
  \begin{eqnarray}
    {\chng\frac{\M_e}{\omega^2}p_{scat}}&=&r(r\ddot{r}+2\dot{r}^2)\nonumber\\
                                       &\approx&-(a+c_1)\cos\tau-(b+D\Omega+c_2)\sin\tau
                                                 +2[2A-a^2+(b+D\Omega)^2]\cos(2\tau)
                                                 +4[B-a(b+D\Omega)]\sin(2\tau)\nonumber\\
                                       &&+\Omega^2\left[D\sin(\Omega\tau)
                                          +2(D^2-2E)\cos(2\Omega\tau)\right]\nonumber\\
                                       &&+\frac{a}{2}\left[
                                          (2D-G-F)(\Omega-1)^2\sin[(\Omega-1)\tau]
                                          +(2D-G+F)(\Omega+1)^2\sin[(\Omega+1)\tau]
                                          \right]\nonumber\\
                                       &&+\frac{1}{2}(b+D\Omega)\left[
                                          (2D-G-F)(\Omega-1)^2\cos[(\Omega-1)\tau)]
                                          -(2D-G+F)(\Omega+1)^2\cos[(\Omega+1)\tau]\right]\,.
                                          \label{pscat}
  \end{eqnarray}
  {\chng To interpret the structure of acoustic bubble response more
    clearly avoiding excessive algebraic detail we consider a particular
    case when noise leads to small expansion of a stationary bubble:
    $0<a\ll1$, $b=0$. In the limit of high-frequency excitation
    $\Omega\gg1$, $F\approx(D/\Omega)\ll D$, $G\approx D$ and
    Eq.~(\ref{pscat}) becomes 
    \begin{eqnarray}
      p_{scat}&\approx&-\omega_0
                        \left[\sqrt{1+\frac{a^2\omega_0^2}{\M_e^2}}
                        \sin(\tau+\phi_4) 
                        +\frac{\K\M_e}{3\omega_0}
                        \sqrt{1+4\frac{a^2\omega_0^2}{\M_e^2}}\sin(2\tau+\phi_5)
                        \right]+\frac{3}{4}\M_e\cos(2\Omega\tau)\nonumber\\
              &&+\sin(\Omega\tau)
                 \left[1+\frac{\M_e}{\omega_0}\sqrt{1+\frac{a^2\omega_0^2}{\M^2_e}}
                 \sin(\tau+\phi_6)\right]\,,\label{pscat1}\\
              &&\ds \phi_4=\tan^{-1}\left[\frac{a\omega_0}{\M_e}
                 -\frac{3+4\K}{12}\frac{\M_e}{\omega_0}\right]\,,\quad
                 \phi_5=\tan^{-1}\left[\frac{a\omega_0}{2\M_e}
                 -\frac{\M_e}{2a\omega_0}\right]\,,
                 \quad \phi_6=\tan^{-1}\frac{a\omega_0}{\M_e}\,.
    \end{eqnarray}
    For $a\lesssim0.1$ the values of $p_{scat}$ obtained using
    Eq.~(\ref{pscat1}) coincide with those computed using a
    numerical solution of the full Eq.~(\ref{NDRPE}) within a few per cent.
    The term involving driving signal $\sin(\Omega\tau)$ in
    Eq.~(\ref{pscat1}) represents an acoustic bubble response after it is
    processed using a combination of low- and high-pass filters as is
    done in our experiments. It defines a classical beating
    pattern {\chng that arises on the background of sinusoidal
    oscillations. (Note that in the absence of random noise ($a=b=0$)
    the relative depth of the resulting beating modulation given by
    $\M_e/\omega_0$ is small. However, the condition $a=b=0$ cannot be 
    fulfilled in the case of a bubble cluster, which means that the 
    beating pattern observed in experiments involving a bubble cluster
    can be much stronger than in the case of a single gas bubble.)} 
    
    We also note that Eq.~(\ref{pscat1}) is a truncated version of the 
    full expression for the acoustic signal scattered by the bubble 
    that only includes terms up to the second harmonics $(2\tau)$ and
    $(2\Omega\tau)$. Yet it captures all main features of the acoustic
    signal structure and shows how AFC is generated around the driving
    ultrasound frequency $\Omega$. The mechanism of generating AFC
    around higher harmonics $n\Omega$, $n>1$ is similar and does not
    need to be discussed separately. We only note that it follows from
    Eq.~(\ref{pscat1}) that the intensity of spectral peaks of
    the acoustic response detected at frequencies $n\Omega$ decreases
    compared to that at $\Omega$ as $\M_e^{n-1}$ and the intensity of
    the $n$th sideband peak is proportional to $(a^n,b^n)$. In
    practical terms this means that the overall AFC width determined by
    the highest detectable ultraharmonic frequency $n\Omega$ can be
    expanded by increasing the amplitude of driving ultrasound waves.}
}
\begin{figure*}[t]
  \centerline{
    \includegraphics[width=14cm]{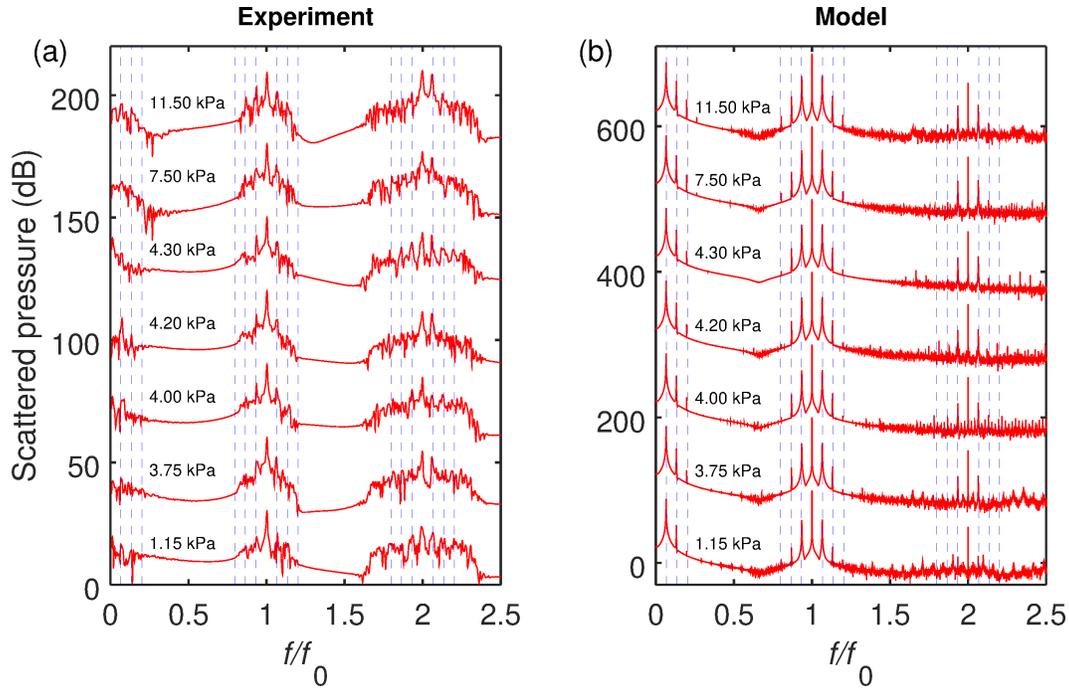}}
  \caption{(a)~Experimental spectra of a cluster of gas bubbles
    in water insonated with the 24.6\,kHz sinusoidal signal of
    increasing pressure amplitude $\alpha=$1.15, 3.75, 4, 4.2, 4.3,
    7.5 and 11.5\,kPa. The frequency axis is normalised with frequency
    $f_0$ of the driving field. {\chng The detectable response frequency
      resolution is $\Delta f/f_0=1.34 \times 10^{-4}$}. The scattered
    pressure values (in dB) are shown along the vertical axis with the
    vertical offset of 30\,dB between spectra. (b)~Calculated spectra
    of a single gas bubble with 1.95\,mm radius at the same driving
    pressure frequency and amplitudes as in the experiment. The
    vertical offset between individual spectra  is 100\,dB. In both
    panels, the vertical dashed lines mark the peaks at the natural
    frequency and its ultraharmonics (the left parts of the spectra)
    as well as the frequencies of the sideband peaks around the
    fundamental and second harmonic frequency of the driving signal.}
  \label{fig:fig2}
\end{figure*}

\section*{Results}
We demonstrate experimentally fundamental physics behind the
principle of the AFC generation, which we suggest, by studying a
cluster of bubbles created using a bubble generator. The natural
frequency of such a cluster is smaller than that of constituent
bubbles because the cluster effectively behaves as a single bubble of
radius $R_c > R_0$ [see Eq.~(\ref{eq:eq6})]. This physical similarity
also enables us to explain experimental findings by conducting
numerical modelling of nonlinear oscillations of a single equivalent
spherical bubble in water, see Methods.

Figure~\ref{fig:fig2}(a) shows the measured dependence of the
scattering spectrum on the increasing amplitude of the driving
pressure $\alpha$ at the driving frequency $f_0=24.6$\,kHz.
We observe the main features of a frequency comb---a number of
equidistant sideband peaks around the fundamental harmonic
frequency $f/f_0=1$ marked by the dashed lines. The distance between
all sideband peaks is 1.67\,kHz, which according to
Eq.~(\ref{eq:eq6}) corresponds to the natural frequency of a 
bubble cluster with $\sqrt{\chi}R_c=1.95$\,mm. The peaks at this
natural frequency and its higher-order ultraharmonics are also 
distinguishable and they are marked by the leftmost dashed line in
Fig.~\ref{fig:fig2}(a). 

We also observe that the nonlinearly induced higher-order
ultraharmonics of the natural response of the bubble cluster 
result in the secondary sideband peaks around $f/f_0=1$.
In fact, the sideband peaks adjacent to the central peak originate
from the natural frequency of the bubble cluster, but the other
two are due to the interference with the first ultraharmonic of the
cluster response at the twice the natural frequency. A qualitatively
similar sideband peak structure can be seen around $f/f_0=2$.

Figure~\ref{fig:fig2}(b) shows the calculated spectra obtained
for the experimental values of the frequency and amplitude.
Consistently with the size of the bubble cluster inferred from the
experiment, in the calculation we assume that the radius of the
single equivalent gas bubble is 1.95\,mm (significantly,
the calculated spectra are qualitatively similar
for the bubble radii in the 1--2\,mm range). We note an overall good
qualitative agreement between the experimental and calculated
spectra. In experiments, we can clearly see the primary and secondary
sidebands around $f/f_0=1$. The calculation also predicts the
existence of the tertiary sidebands at high values of
$\alpha$. However, these are undetectable in our measurements due to
their low relative magnitude.

{\chng The spectra of the experimental response contain components
  that are not found in calculations even after we remove the random
  noise floor from raw measured traces. As discussed in section
  Methods we exclude the possibility of a significant external noise
  due to vibrations and interference with high-voltage equipment. Thus
  we relate a slight irregularity of the comb lines to the Doppler effect
  associated with a translational motion of oscillating bubbles in
  the incident ultrasound field \cite{Pou16, Lee16}. The size variation 
  of the generated bubbles could also contribute to the comb line
  imperfection. However, we do not consider these deficiencies of
  initial proof-of-concept experiments as prohibitive for the
  following reasons. The position of a single oscillating bubble can
  be stabilised by trapping it in the antinode of an acoustic standing
  wave field in a resonator \cite{Put00, Gei00, Bre02}. Similar
  techniques can be used to stabilise a cluster of bubbles
  \cite{Pen19}, which would significantly reduce the random Doppler
  distortion of acoustic comb spectra. The technical problem with a
  bubble size variations can also be resolved by using existing higher
  precision mono-disperse bubble generators (albeit more expensive
  than that currently available to us) \cite{Dol08, Che09}.}

Interestingly, the tertiary sideband peaks can be seen around
$f/f_0=1$ at $\alpha=4.3$\,kPa. More broadly, we note a larger 
magnitude of all experimental peaks at $f/f_0=2$ compared
to the calculated values. This observation is consistent with the fact
that a response of a bubble cluster rather than of a single bubble is
measured: clusters exhibit stronger acoustic nonlinearities
\cite{Rud06, Ain11} that give rise to more energetic signals at the
second harmonic frequency $f/f_0=2$. 
\begin{figure*}[t]
  \centerline{\includegraphics[width=12cm]{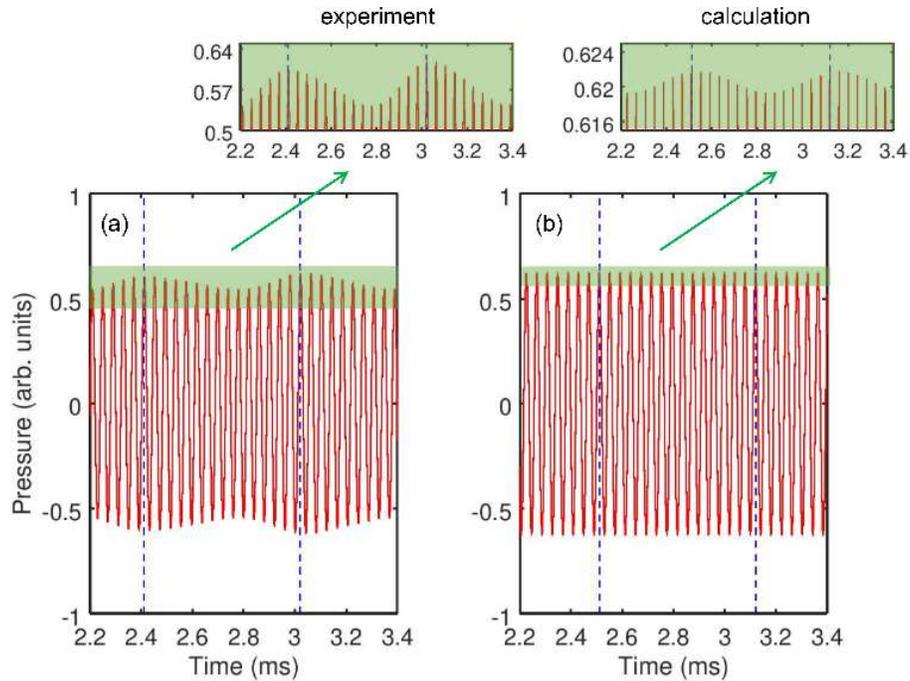}}
  \caption{(a)~Measured acoustic response of the gas bubble cluster and
  (b)~calculated acoustic response of a single equivalent bubble
  corresponding to a sinusoidal pressure wave with the frequency
  $f_0=24.6$\,kHz and amplitude $\alpha=11.5$\,kPa. {\chng In the
  calculation, the initial conditions were set to be $R(0)=R_0$ and 
  $\frac{d}{dt}R(0)=0$, see Eq.~(\ref{eq:eq1})}. The time between 
  the vertical dashed lines is $\Delta T=1/f_{nat} \approx 0.6$\,ms. The 
  insets show the closeup of the waveforms and demonstrate
  the amplitude modulation \chng{(see the main text for more details.)}
  Arbitrary pressure units are used in both panels to enable the 
  comparison of the experimental and calculated data.}
  \label{fig:fig3}
\end{figure*}

Figure~\ref{fig:fig3}(a) shows the temporal far-field pressure
profile corresponding to $f_0=24.6$\,kHz (i.e.~$f/f_0=1$)
at $\alpha=11.5$\,kPa, [the top spectrum in
Fig.~\ref{fig:fig2}(a)]. {\chng These profiles were obtained
  by digitally filtering the measured signal using a narrow
  bandpass filter that was designed so that the central
  frequency would correspond to 24.6\,kHz but the filter
  bandwidth would be wide enough not to cut the sidebands.}
We observe an amplitude-modulated signal with
the modulation period close to that of the natural bubble 
cluster oscillations. 

{\chng In the temporal profile in Fig.~\ref{fig:fig3}(a), the amplitude
  modulation depth (the ratio of the modulation excursions 
  to the amplitude of the unmodulated carrier wave) is smaller than 1.
  In some optical comb technologies, where direct photodetection 
  of the optical pulses is used to produce an electronic
  signal that follows the amplitude modulation of the pulse train,
  low modulation depth could pose challenges for practical realisation 
  \cite{Pic19}. However, this does not present a problem
  in our case because the frequencies of the electronic signal 
  of AFC coincide with the frequencies of the driving pressure wave, 
  which dramatically simplifies the characterisation of the comb.}

The experimental time series is in good qualitative agreement with the
calculated one shown in Fig.~\ref{fig:fig3}(b). According to the
discussion in Methods, the ultrasound energy absorption and scattering
by the equivalent bubble are stronger than those by the bubble
cluster. Because in the calculation the power of the driving
ultrasound wave is the same as in the experiment, the oscillations of
the equivalent bubble at its natural frequency are less
energetic. Therefore, their interaction with the driving ultrasound
waves should result in a weaker amplitude modulation.
{\chng On the other hand, our asymptotic analysis of Eq.~(\ref{eq:eq1})
demonstrates that in our model of the equivalent bubble the amplitude
modulation also depends on the chosen initial conditions. Indeed, by
varying in the calculation the initial radius of the bubble within $\pm$5\%
we could achieve a nearly the same amplitude modulation as 
in the experiment. However, because of a complex interaction
between the bubbles within the cluster \cite{Doi01, Nas13, Mae19}
and therefore essentially random initial conditions, it is challenging
to establish the actual physical mechanism responsible for
the discrepancy in the amplitude modulation in Fig.~\ref{fig:fig3}(a)
and Fig.~\ref{fig:fig3}(b).}

\begin{figure}[t]
  \centerline{\includegraphics[width=8.3cm]{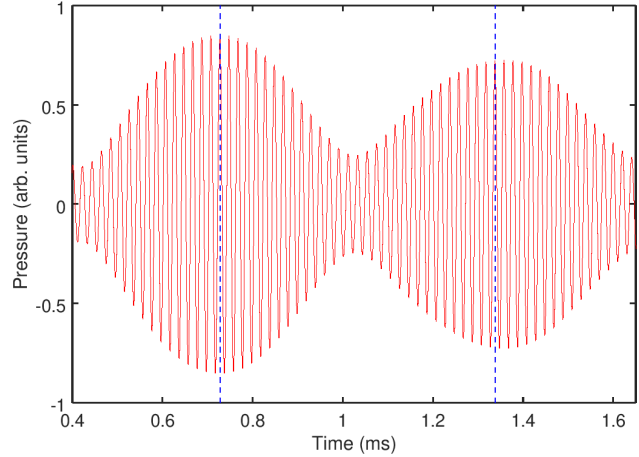}}
  \caption{(a)~Measured acoustic bubble cluster response 
    corresponding to a sinusoidal driving pressure wave with the frequency
    $f_0=49.2$\,kHz {\chng (twice the frequency in 
      Fig.~\ref{fig:fig3})} and amplitude $\alpha=4.3$\,kPa. The time between 
    the vertical dashed lines is $\Delta T=1/f_{nat} \approx 0.6$\,ms.}
  \label{fig:fig4}
\end{figure}

Next we focus on the experimental sideband peak structures at 
$f_0=49.2$\,kHz (i.e.~$f/f_0=2$) at $\alpha = 4.3$\,kPa 
because it has three sidebands on each side. As shown in
Fig.~\ref{fig:fig4}, the amplitude modulation gives rise
to a train of pulses with the modulation period close
to that of the natural bubble cluster oscillations.
     
{\chng We also note a stronger irregularity of the envelope shape in
  Fig.~\ref{fig:fig4} compared with the profile in
  Fig.~\ref{fig:fig3}(a), where a slight irregularity can only be seen
  in the close-up [inset in Fig.~\ref{fig:fig3}(a)]. As discussed
  above, we attribute this difference to a translational motion of
  bubbles in the cluster \cite{Doi01, Nas13, Mae19}. The analysis of
  such a translation motion poses significant computational and
  experimental challenges \cite{Mae19}. However, a theoretical insight
  can be gained from the analysis of a system consisting of just two
  moving gas bubbles \cite{Doi01}, where it has been shown that
  stronger ultrasound forcing prevents bubble collision and
  coalescence thereby increasing the stability of their
  oscillations. On the other hand, similar to a single oscillating
  bubble \cite{Wat93, Doi02, Mat03, Met09, Dey16}, the translational
  motion in bubble clusters should depend on the frequency and
  pressure of the driving wave. Thus, it is plausible to assume that
  driving the bubble cluster at a higher frequency but lower pressure,
  which is the case in Fig.~\ref{fig:fig4}, could compromise the
  stability of the cluster, which indeed has been predicted in
  ref.~\onlinecite{Nas13}. Clearly, any deterioration of the cluster
  stability would adversely affect the coherence of its oscillations,
  which we believe is the reason for an imperfection of the envelope
  shape in Fig.~\ref{fig:fig4}. Admittedly, this effect is
  unfavourable for the generation of a frequency comb, and its further
  studies would be important from both the fundamental and practical
  points of view.} 

Finally, we note that the spectral peak structure of the AFCs
demonstrated in this work should be stable with respect to the changes
in the viscosity, density and surface tension of water due to the
variations in the temperature and other environmental factors such as
the salinity of water. Indeed, whereas the material parameters of
water depend on ambient conditions and affect the natural bubble
frequency in principle [see Eq.~(\ref{eq:eq5})], such an influence on
mm-size bubbles is negligible as discussed in the Model section.
 
\section*{Discussion}
AFC technique is an emerging metrological approach that benefits from
technological maturity of optical frequency combs. It opens up
opportunities for accurate measurements in various  physical, chemical
and biological systems in situations, where using light poses
technical and fundamental limitations, for example, when precise
underwater distance measurement is required \cite{Wu19}.

{\chng Whereas AFCs are expected to operate similarly to optical combs,
  there are a number of differences between these two techniques
  associated with the disparate frequencies of acoustic waves and
  light and mechanisms of the interaction of these waves with the
  medium \cite{Mak19}. This aspect presents numerous technological 
  challenges that shape research efforts in {\chng the field
    \cite{Pic19, For19}. We also note that frequency combs with a flat
    spectral intensity distribution are preferred in a number of
    applications \cite{Pic19, For19}. In AFCs this has yet to be
    achieved. Moreover,} as shown in refs\,\cite{Mak19, Gan17}, AFCs
  have the smaller number of spectral lines compared with optical
  combs. This feature has been identified as being important for a
  number of practical applications including phonon lasers
  \cite{Gru10, Bea10} and computing \cite{Sta12}. 

  In view of the above, AFCs should be compared with either Brillouin
  \cite{Mak17, Gar18} or opto-electronic combs \cite{Tor14} that are
  known to also have a small number of lines and require special
  techniques for broadening their spectral ranges \cite{Zha19_1}. The
  same applies, although to a lesser degree, to Kerr optical combs that are
  generated using a cascade of optical four-wave mixing processes in a
  photonic microresonator \cite{Che16}. Kerr combs may have just
  ten or so lines due to intrinsically low strength of nonlinear-optical 
  effects \cite{Mak13, Mak19}, which is in stark contrast to
  conventional optical combs based on mode-locked lasers and
  consisting of hundreds of lines \cite{Die96}.} 

In good agreement with our numerical predictions, our experimental
results demonstrate that a signal produced by gas bubbles oscillating
in water has a frequency spectrum composed of equidistant peaks and is
characterised by amplitude modulation at the bubble cluster
natural frequency. These features are similar to those of typical optical 
frequency combs and thus they demonstrate the feasibility of the
AFC generation by using gas bubble oscillations in a liquid.
{\chng Although we were able to generate just a few comb lines,
  we found that their structure was similar to that produced in earlier 
  works \cite{Gan17}. It should suffice for such practical
  applications as phonon lasers \cite{Gru10, Bea10} for the
  realisation of which an acoustic resonator filled with a liquid
  containing gas bubbles acting as the active medium was suggested
  \cite{Zav96}. In such a scheme, bubble oscillations have to be
  self-synchronised to resonate at a certain frequency, which can be
  achieved by appropriately tuning the AFC lines. Moreover, some
  techniques for spectral broadening of optical combs based on
  nonlinear-optical effects \cite{Tor14} can be applied to increase
  the number of lines in a comb based on gas bubble oscillations
  because of the analogy between nonlinear optical and 
  acoustical processes \cite{Mak19}.} {\chng For example, as 
  follows from our asymptotic analysis of Eq.~(\ref{eq:eq1}), 
  the number of lines in our AFC and their relative magnitude 
  can be increased by simultaneously decreasing the frequency 
  and increasing the pressure of the forcing ultrasound wave. 
  As a result, one can obtain a spectrum where the sideband peaks 
  around the fundamental and second harmonic frequency of the 
  driving signal form a continuous comb structure. However, one should bear
  in mind that the frequency of the forcing has always to be 
  higher than the natural oscillation frequency of the bubble cluster.}

The so-generated AFCs should {\chng also} find an application niche
in the fields of underwater distance measurements and
communication. However, their wider use is expected to be in the areas
of biology and medicine, where there is a need for novel types of
biomedical sensors. For example, the AFCs suggested here can be used to
measure elastic properties of some biological tissues and living cells and
sensing biochemical processes inside them via inducing elastic
deformation in the proximity of an oscillating bubble \cite{Zin05,
Jal20}. Such a local mechanical deformation would affect
the oscillation dynamics of the bubble \cite{Doi11} and lead to
detectable modifications of the sideband spectral structure of the
comb. Thus, it should be possible to use bubbles oscillating in water
contaminated with pathogens (e.g.~bacteria) to obtain information
about their presence and concentration {\chng that are required for
  choosing an adequate strategy for their removal \cite{Vya20} or
  disinfection \cite{Zin09, Boy20}.}

Our AFCs can also be used to measure the resonance frequency of a
bubble of unknown size \cite{Lei91, Lei96}. Thus far, a number of
bubble sizing techniques using two-frequency excitation have been
employed \cite{Lei91, Lei96}. There, two beams---a pump beam of
variable frequency and an imaging beam of fixed frequency---are
simultaneously used to scan across the expected resonance
frequency of the bubble and to achieve the coupling between
the two signal, when the bubble undergoes nonlinear oscillations
at resonance. Using a frequency comb generated with just one driving
wave will extend the capability of this technique because,
from the technical point of view, only one ultrasound transducer
needs to be employed.

\section*{Data Availability}
The datasets generated during and/or analysed during the current
study are available from the corresponding author on reasonable request.

\providecommand{\noopsort}[1]{}\providecommand{\singleletter}[1]{#1}%

\section*{Acknowledgements}

This work was supported by the Australian Research Council
(ARC) through the Future Fellowship (FT180100343) program
and by Swinburne University through the Swinburne
University Postgraduate Research Award (SUPRA) scheme. 

\section*{Author contributions}

{\chng B.Q.H.N.~developed the computational and symbolic algebra codes for
  solving the Rayleigh-Plesset equation. I.S.M.~conceived the idea and 
  conducted the experiment. S.A.S.~developed an asymptotic analysis of 
  the bubble oscillation model. I.S.M.~and S.A.S.~analysed and interpreted 
  the results and wrote the manuscript. All authors contributed to editing
  and revising the manuscript.}
\\
\noindent\textbf{Competing interests:} The authors declare no
competing interests.

\end{document}